\documentclass[aps,prl,twocolumn,showpacs]{revtex4}
\newcommand{\be}{\begin{equation}}
\newcommand{\ee}{\end{equation}}
\newcommand{\ben}{\begin{eqnarray}}
\newcommand{\een}{\end{eqnarray}}
\usepackage{graphicx,epsfig}
\usepackage{amsfonts}
\usepackage{amssymb}
\begin{document}
\title{New Global Defect Structures}

\author{D. Bazeia, J. Menezes, and R. Menezes}
\affiliation{Departamento de F\'\i sica, Universidade Federal da Para\'\i ba,
C.P. 5008, 58051-970 Jo\~ao Pessoa PB, Brazil}

\date{\today}
\begin{abstract}
We investigate the presence of defects in systems described by real scalar
field in (D,1) spacetime dimensions. We show that when the potential assumes
specific form, there are models which support stable global defects for D
arbitrary. We also show how to find first-order differential equations
that solve the equations of motion, and how to solve models in D
dimensions via soluble problems in D=1. We illustrate the procedure
examining specific models and finding explicit solutions.
\end{abstract}
\maketitle

The search for defect structures of topological nature is of direct interest
to high energy physics, in particular to gravity in warped spacetimes involving
$D$ spatial extra dimensions. Very recently, a great deal of attention has
been given to scalar fields coupled to gravity in $(4,1)$
dimensions \cite{rs,gw,gt,grs,df}. Our interest here is related to
Ref.~{\cite{campos}}, which deals with critical behavior of thick branes
induced at high temperature, and to Refs.~{\cite{G,GS,RS}}, which study the
coupling of scalar and other fields to gravity in warped spacetimes involving
two or more extra dimensions.

These specific investigations have motivated us to study defect solutions
in models involving scalar field in $(D,1)$ spacetime dimensions. To do this,
however, we have to circumvent a theorem \cite{H,D,J}, which states that 
models described by a single real scalar field cannot support topological
defects, unless we work in $(1,1)$ space-time dimensions. To evade this
problem, in the present letter we consider models described by the Lagrange
density $\mathcal{L}=(1/2)\partial_{\mu}\phi\partial^{\mu}\phi-U(x^2; \phi),$
where the potential is $U(x^2;\phi)=f(x^2){V(\phi)},$ $x^2=x_{\mu}x^{\mu}$,
and $\phi$ is a real scalar field. The metric is $(+,-,\ldots,-)$, with
$x_\mu=(x_0,x_1,x_2,...,x_D)$ in $(D,1)$ space-time dimensions.
$V(\phi)$ has the form $V(\phi)=W^2_\phi/2$, where $W=W(\phi)$
is a smooth function of $\phi$, and $W_{\phi}=dW/d\phi$.
We suppose that ${\bar\phi}$ is a critical point of $V$, such that
$V({\bar\phi})=0$. This generalization is different from the
extensions one usually considers to evade the aforementined problem,
which include for instance constraints in the scalar fields and/or the
presence of fields with nonzero spin -- see, e.g., Ref.~{\cite{vs}}, and 
other specific works on the subject \cite{ddi,afz}. Potentials of the above
form appear for instance in the Gross-Pitaevski equation, which finds
applications in several branches of physics -- see, e.g., Ref.~{\cite{GP}}.
Other recent examples in $(1,1)$ dimensions include Ref.~{\cite{SV}}, which
deals with the dynamics of embedded kinks, and Refs.~{\cite{OG}}, which
describe scalar field in distinct backgrounds.

In higher dimensions, the factor $1/r^N$ that we
introduce in Eq.~(\ref{pot}) gives rise to an effective model, which
comes from a more fundamental theory. To make this point clear,
we consider the model
${\cal L}_1=\partial_\mu\phi\partial^\mu\phi-f(\phi)F_{\mu\nu}F^{\mu\nu},$
which is a simplified Abelian version of the color dielectric model \cite{fl}
in the absence of fermions -- see, e.g., Ref.~{\cite{wil}}. This model
describes coupling between the real scalar field and the gauge field
$A_\mu$, through the dielectric function $f(\phi).$
Here $F_{\mu\nu}=\partial_\mu A_\nu-\partial_\nu A_\mu$ is the gauge field
strenght. This model shows that for spherically symmetric static
configurations in the electric sector, the equation of motion for
the matter field is $\nabla\phi+(df/d\phi)E_r^2=0,$ where
${\vec E}=(E_r,0,...,0)$ is the electric field.
The use of the equation of motion for the electric field
provides an effective description of the matter field $\phi,$ in which
the equation of motion exactly reproduces \cite{bmm} the equation of motion
that we will be investigating below, under the choice $f(\phi)=1/V(\phi).$
As one knows, color dielectric models may provide effective descriptions
for non-perturbative QCD -- see, e.g., Ref.~{\cite{ws}} for a very recent
investigation and for related works.

We now focus attention on the effective model. We investigate the presence
of static solutions considering the specific potential
\be\label{pot}
U(x^2;\phi)=\frac1{2r^N} W^2_\phi
\ee
where $r=(x_1^2+x_2^2+\cdots+x_D^2)^{1/2}$. We write $\phi=\phi(\vec{r})$
to obtain the equation of motion
$\nabla^2\phi=(1/2r^N)\,W_\phi W_{\phi\phi}.$ The energy density of the
field configuration has the form
${\mathcal E}=(1/2)(\nabla\phi)^2+(1/2r^N)W^2_\phi.$
We suppose that $\phi=\phi(\vec{r})$ solves
the equation of motion. It has total energy
$E^D=E_g^D+E_p^D$, which splits into gradient and potential portions.
We make the change
$\phi(\vec{r})\to\phi^\lambda(\vec{r})= \phi(\lambda\vec{r})$
to get to the conditions
\be
(2-D)\,E_g^D+(N-D)\,E_p^D=0 \label{rel1}
\ee
and $(2-D)(1-D) \,E_g^D+(N-D)(N-D-1)\,E_p^D\geq0,$
to make the field configuration stable. Since $E_g^D\geq0$
and $E_p^D\geq0$, these conditions impose restrictions on both
$N$ and $D$.

An important case is $D=1$, and here
the value $N=0$ makes $E_g^1=E_p^1$, that is, the energy
is equally shared between gradient and potential portions. This is the
standard result: from \cite{H,D,J} we see that for the special
case $N=0$, there exist stable solutions only in $D=1$. However,
the form of the potential (\ref{pot}) is peculiar, and it allows obtaining
several other cases which support defect solutions. In two spatial dimensions,
for $D=2$ one gets $N=2$, but this gives no further relation between the
gradient and potential portions of the energy. For $D\geq3$ we get that
$N=2D-2$ to make $E_g^D=E_p^D$.

We investigate the possibility of obtaining a Bogomol'nyi bound \cite{bog,ps}
for the energy of static configurations in the present investigation.
The case $D=1$ is standard, so we deal with $D\geq2$. We suppose that
the static solutions engender spherical symmetry, that is, we consider
$\phi=\phi(r)$, with $r\in[0,\infty)$. The energy can be written 
as $E^D\geq\Omega_D |\Delta W|$, where
$\Delta W = \pm W[\{\phi(r\to\infty)\}]\mp W[\{ \phi(r=0) \}]$
and $\Omega_D$ is the D-dependent angular factor. This result is
obtained if and only if $N=2D-2$, and the energy is minimized to
$E^D=\Omega_D|\Delta W|$ for static and radial field configurations
which solve the first order equations
\be\label{foeq}
\frac{d\phi}{dr}=\pm \frac1{r^{D-1}}W_{\phi}
\ee
This is the Bogomol'nyi bound, now generalized to models for scalar field
that live in $D$ spatial dimensions. The value $N=2D-2$ leads to defect
solutions of the BPS type, which obey first order equations and have energy
evenly split into gradient and potential portions. For $D=2$ one gets
$N=2$, and in this case the model engenders scale invariance. We can show
that solutions of the above first order equations solve the equation of
motion (\ref{em}) for potentials given by Eq.~(\ref{pot}). Also, we follow
Ref.~{\cite{bms}} and introduce the ratio $R=r^{D-1}(d\phi/dr)/W_\phi$.
For field configuration that obey $\phi(0)=\bar{\phi}$ and
$\lim_{r\to0}d\phi/dr\to0$ one can show that solutions of the equation
of motion also solve the first order equations (\ref{foeq}). This extends
the result of Ref.~{\cite{bms}} to the present investigation: it shows that the
equation of motion (\ref{em}) completely factorizes into the two first
order equations (\ref{foeq}).

The equation of motion for $\phi=\phi(r)$ is
\be\label{em}
\frac1{r^{D-1}}\frac{d}{dr}
\left(r^{D-1}\frac{d\phi}{dr}\right)=\frac1{r^{2D-2}}W_\phi W_{\phi\phi}
\ee
The first order equation is given by (\ref{foeq}), and their solutions solve
the equation of motion and are stable against radial, time-dependent
fluctuations. To see this we consider
$\phi(r,t)=\phi(r)+\sum_k \eta_k(r) \cos(w_{k}\, t)$. For small fluctuations
we get $H\eta_k=w^2_{k}\eta_k$, where the Hamiltonian can be written as
\be
H=\frac{1}{r^{2D-2}}\left(-r^{D-1}\frac{d}{dr}\mp W_{\phi\phi}\right)
\left(r^{D-1}\frac{d}{dr}\mp W_{\phi\phi}\right)
\ee
It is non-negative, and the lowest bound state is the zero mode, which obeys
$r^{D-1}d\eta_0/dr=\pm\,W_{\phi\phi}\,\eta_0.$ This gives
$\eta_0(r)=c\,\exp\left({\pm\int dr \,{r^{1-D}}\,W_{\phi\phi}}\right),$
where $c$ is the normalization constant, which usually exists only for one of
the two sign possibilities. We can also write $\eta_0(r)=c W_\phi$, which
is another way to write the zero mode.

We investigate the presence of defect structures turning attention
to specific models in $D=1,\;D=2,$ and $D\geq3$. We first consider
the case $D=1.$ We choose $N=0$. The equation of motion is
$(d^2\phi/dx^2)=W_\phi\,W_{\phi\phi}$. We are searching for solutions
that obey the boundary conditions $\lim_{x\to-\infty}\phi(x)\to{\bar{\phi}}$
and $\lim_{x\to\-\infty}(d\phi/dx)\to0$, where $\bar{\phi}$ is a
critical point of the potential, obeying $V(\bar\phi)=0$. In this case
one can write $(d\phi/dx)=\pm(dW/d\phi);$ see \cite{bms}.
In $D=1$ one usually introduces the conserved current
$j^\mu=\varepsilon^{\mu\nu}\partial_\nu\phi$. We see that $\rho=d\phi/dx$
and so $\rho^2$ gives the energy density of the field configuration \cite{bb}.
Thus, we introduce $Q_T=\int^{\infty}_{-\infty}dx \rho^2$ as  the topological
charge, which is exactly the total energy of the solution.

We exemplify the case $D=1$ with the family of models
\be\label{model}
V_p(\phi)=\frac12\,\phi^{2}
\left(\phi^{\frac{-1}{p}}-\phi^{\frac{1}{p}}\right)^2
\ee
The parameter $p$ is real, and it is related to the way the field
self-interacts. These models are well-defined for $p$ odd, $p=1,3,5...$
For $p=1$ we get to the standard $\phi^4$ theory. For $p=3,5,...$ we have
new models, presenting potentials which support minima at $\bar\phi=0$
and $\pm1$. For $p$ odd the classical bosonic masses at the asymmetric minima
$\bar\phi=\pm1$ are given by $m^2=4/p^2$. For $p=3,5,...$ another minimum
appear at $\bar\phi=0$. However, the classical mass at this symmetric
minimum diverges, signalling that $\bar\phi=0$ does not define a true
perturbative ground state for the system.

We consider $p$ odd. The first-order equations are
$d\phi/dx=\pm\phi^{\frac{p-1}{p}}\mp\phi^{\frac{p+1}{p}},$
which have solutions $\phi^{(1,p)}_{\pm}(x)=\pm\tanh^p(x/p).$
We consider the center of the defect at ${\bar x}=0$, for simplicity.
Their energies are given by $E^{(p)}_{1,o}=4p/(4p^2-1)$, and we plot
some of them in Fig.~[1].

\begin{figure}[h]
\includegraphics[{height=2.7cm,width=7.5cm}]{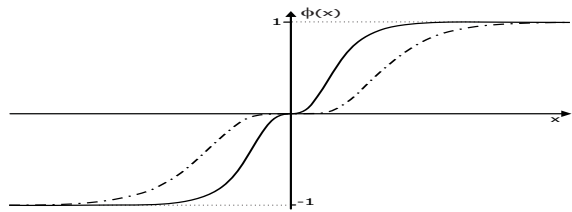}
\caption{Defect solutions for models with $p=3\;{\rm and}\;5,$
respectively full and dashed-dotted lines.}
\end{figure}

We see that solutions for $p=3,5,...,$ connect the minima
$\bar{\phi}=\pm1$, passing through the symmetric minimum
at $\bar{\phi}=0$ with vanishing derivative. They are new structures, which
solve first-order equations, and we call them 2-kink defects since they seem
to be composed of two standard kinks, symmetrically separated by a distance
which is proportional to $p$, the parameter that specifies the potential.
To see this, we notice that the zero modes are given by
$\eta_0^{(1,p)}=c^{(1,p)}\tanh^{p-1}(x/p)\,{\rm sech}^2(x/p),$
where $c^{(1,p)}=[(4p^2-1)/4p]^{1/2}$ -- we use $\eta_0^{(D,p)}$ to represent
the zero modes. These zero modes concentrate around two symmetric points,
which identify each one of the two standard kinks. Defect structures similar
to the above 2-kink defects have been studied in the recent past, for instance
in Ref.~{\cite{chw}}, for solutions of the equations of motion in a
$Z(3)$-symmetric model, and also in Ref.~{\cite{svo}}, in a supersymmetric
theory, for solutions which solve first-order equations. We have
studied the $V_p(\phi)$ model coupled to gravity in warped spacetimes, in
$(4,1)$ dimensions \cite{bfg}. The results show that it also induces critical
behavior similar to Ref.~{\cite{campos}}, but now the driving parameter is $p$,
which indicates the way the scalar field self-interacts, and not the
temperature anymore.

The cases with $p$ even are different. The case $p=2$ is special: it gives
the potential $V(\phi)=\phi/2-\phi^2+\phi^3/2$, which supports the
nontopological or lumplike solution $\phi^{(1,2)}_l(x)=\tanh^2(x/2).$
The lumplike solution is unstable, as we can see from the zero mode,
which is proportional to $\tanh(x/2){\rm sech}^2(x/2)$: the zero mode has a
node, so there must be a lower (negative energy) eigenvalue. In the present
case, the tachyonic eigenvalue can be calculated exactly since the
quantum-mechanical potential associated with stability of the lumplike
configuration has the form $U(x)=1-3\, {\rm sech}^2(x/2)$.
This potential supports three bound states, the first being
a tachyonic eigenfunction with eigenvalue $w^2_0=-5/4$, the second
the zero mode, $w^2_1=0,$ and the third a positive energy bound state
with $w^2_2=3/4$. As one knows, tachyons appear in String Theory \cite{W,KS}
and this has brought renewed interest on the subject -- see, e.g.,
Refs.~{\cite{S,Z,MZ}}.

The other cases for $p$ even are $p=4,6,...$ These cases require that
$\phi\geq0$ in Eq.~(\ref{model}), but we can also change $\phi\to-\phi$
in Eq.~(\ref{model}) and consider $\phi\leq0$. We investigate the case
$\phi\in[0,\infty)$; reflection symmetry leads to the other case.
We notice that the origin is also a minimum, with null derivative.
These models also support topological defects, in the form
$\phi^{(1,p)}(x)=\tanh^p(x/p)\;(x\geq0)$, with energies
$E^{(p)}_{1,e}=2p/(4p^2-1)$ for $p=4,6,...$ These solutions solve
the first order equation $d\phi/dx=W_\phi$; the other equation
$d\phi/dx=-W_\phi$ is solved by $\phi(x)=-\tanh^p(x/p)\;(x\leq0)$.

We now consider $D=2$, and $N=2$.
The equation of motion is $\nabla^2\phi=(1/r^2)W_\phi W_{\phi\phi}$.
We search for solution $\phi(r)$, which only depends on the radial coordinate,
obeying $\phi(0)=\bar{\phi}$ and $\lim_{r\to0}d\phi/dr\to0$.
In this case we get
\be
r\frac{d}{dr}\left(r\frac{d\phi}{dr}\right)=
\frac{dW}{d\phi}\frac{d^2W}{d\phi^2}
\ee 
We write $dx=\pm r^{-1}dr$ to get $d^2\phi/dx^2=W_\phi W_{\phi\phi}.$
This result maps the $D=2$ model into the $D=1$ model.

We see that $r=\exp(\pm x),$ which shows that the full line
$x\in(-\infty,\infty)$ is mapped to $r\in[0,\infty)$. We notice
that since the center of the defect is arbitrary, so is the point $r=1$
in $D=2$; thus the solution introduces no fundamental scale,
in accordance with the scale symmetry that the model engenders. If one uses
the model (\ref{model}) with $p$ odd to define the potential in this case,
we get the solutions
\be
\phi_{\pm}^{(2,p)}(r)=
\pm\left(\frac{r^{2/p}-1}{r^{2/p}+1}\right)^p
\ee
Their energies are $E^{(p)}_2={8\pi p}/(4p^2-1)$. In Fig.~[2] we depict
the defect solution for $p=1$. The corresponding zero mode is given by
$\eta^{(2,1)}_0(r)=\sqrt{32/\pi}[r^2/(r^2+1)^2].$
This zero mode binds around the circle where the defect solution vanishes,
as we show in Fig.~[2], where we also plot
$\rho^{(2,1)}_0(r)=32r^4/\pi(r^2+1)^4$.

\begin{figure}[h!]
\includegraphics[{height=3.2cm,width=7.5cm}]{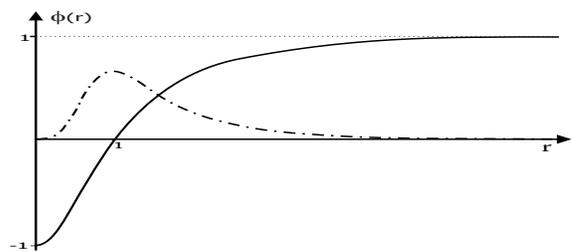}
\caption{Radial defect solutions for $D=2$, in the case $p=1$.
The dashed-dotted line represents the density $\rho^{(2,1)}_0(r)$ of the
corresponding zero mode.}
\end{figure}

The last case is $D\geq3$, with $N=2D-2$.
The equation of motion for $\phi=\phi(r)$ is
\be
r^{D-1} \frac{d}{dr}\left(r^{D-1}
\frac{d\phi}{dr}\right)=W_\phi W_{\phi\phi}
\ee
We write $dx=\pm r^{1-D}dr$ to get $d^2\phi/dx^2=W_\phi W_{\phi\phi}$
and again, we map the model into a one-dimensional problem. We solve
$dx=\pm r^{1-D}dr$ to get $x=\mp r^{(2-D)}/(D-2)$. This shows that
$x$ is now in $(-\infty,0]$ or $[0,\infty)$, and so we have to use
the upper sign for $x\leq0$, or the lower sign for $x\geq0$.
We can use the model of Eq.~(\ref{model}) for $p=4,6,...$ to
solve the D-dimensional problem with $N=2D-2$. In this case the
solutions are
\be
\phi^{(D,p)}(r)=\tanh^{\,p}\biggr[\,\frac1p\,
\left(\frac{r^{2-D}}{D-2}\right)\biggl]
\ee
for $D=3,4,...$ Their energies are given by
$E_D^{(p)}=\Omega_D[{2p}/(4p^2-1)]$, and in Fig.~[3] we depict the
solution for $D=3$.

In the case of $D=3$ and $p=4$ the zero mode is given by
$\eta_0^{(3,4)}(r)=c^{(3,4)}\tanh^3(1/4r){\rm sech}^2(1/4r);$
we could not find the normalization factor explicitly in this case.
We plot $\rho^{(3,4)}_0(r)$ in Fig.~[3] to show that the zero mode
binds at the skin of the defect solution, concentrating
around the radius $R$ of the defect, which is given
by $R=1/4\,{\rm arctanh}[(1/2)^{1/4}]$.

The solutions that we have found have a central core, and a
skin which depends on the parameters that specify the potential of the model.
They are stable, well distinct of other known defects such as the bubbles
formed from unstable domain walls -- see for instance
Refs.~{\cite{b1,b2,b3,b4}}. Also, they are neutral structures,
and may contribute to curve spacetime, and to affect cosmic evolution.
They may become charged if charged bosons and/or fermions bind to them. For
instance, if one couples Dirac fermions with the Yukawa coupling
$Y(\phi)=r^{1-D}\,W_{\phi\phi}$, the fermionic zero modes are similar
to the bosonic zero modes that we have just presented. In Ref.~{\cite{bmm}}
we investigate these and other issues in detail.

\begin{figure}[h!]
\includegraphics[{height=2.5cm,width=7.5cm}]{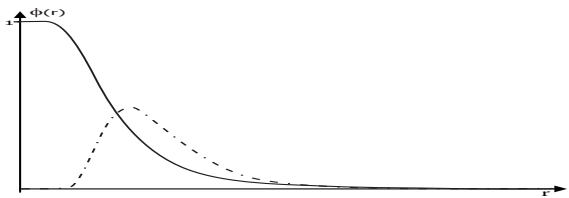}
\caption{Radial defect solutions for $D=3$, in the case $p=4$. The
dashed-dotted line shows (1/10 of) the density of the corresponding
zero mode.}
\end{figure}

We end this letter recalling that we have solved the equation of motion using
$r^{1-D}dr=\pm dx$. Although this identification works very naturally for
the first order equations, we used the equation of motion to present a more
general investigation, which may be extended to tachyons in $D$ spatial
dimensions. For instance, from the lumplike solution for $D=1$ and $p=2$
we can obtain another lumplike solution $\phi^{(2,2)}_l(r)=(r-1)^2/(r+1)^2,$
which is valid for $D=2$. Another issue concerns the identification of the
topological behavior of the above defect structures. We do this introducing
the (generalized current-like) tensor
$j^{\mu_1\mu_2\cdots\mu_{D}}=
\varepsilon^{\mu_1\mu_2\cdots\mu_{D}\mu_{D+1}}\partial_{\mu_{D+1}}\phi.$
It obeys $\partial_{\mu_i}j^{\mu_1\mu_2\cdots\mu_{D}}=0$
for $i=1,2,\cdots,D,$ which means that the quantities
$\rho^{i_1i_2\cdots i_{D-1}}=j^{0i_1i_2\cdots i_{D-1}}$
constitute a family of $D$ distinct conserved (generalized charge)
densities. We introduce the scalar quantity
$\rho_D^2=\rho_{i_1i_2\cdots i_{D-1}}\rho^{i_1i_2\cdots i_{D-1}}=(-1)^D(D-1)!
(d\phi/dr)^2,$
which generalizes the standard result, obtained with
$j^\mu=\varepsilon^{\mu\nu}\partial_\nu\phi$ -- see the reasoning
above Eq.~(\ref{model}). Thus, we define the
topological charge as
$Q^D_T=\int d{\vec r}\,\rho_D^2=(-1)^D(D-1)!\Omega_D\Delta W$,
which exposes the topological behavior of the new global defect
structures that we have just found.

We would like to thank R. Jackiw for drawing our attention to
Ref.~{\cite{H}}. We also thank CAPES, CNPq, PROCAD and PRONEX
for partial support.


\end{document}